%% file: ISMIR2026_template.tex
\documentclass{article}
\usepackage[T1]{fontenc}
\usepackage[utf8]{inputenc}
\usepackage{ismir}
\usepackage{amsmath,amssymb,cite,url}
\usepackage{graphicx}
\usepackage[dvipsnames]{xcolor}
\usepackage{tikz}
\usetikzlibrary{positioning, shapes.geometric, arrows.meta, fit, backgrounds, calc, decorations.pathmorphing}
\usepackage{booktabs}
\usepackage{colortbl}
\usepackage{multirow}
\usepackage{subcaption}

\newcommand{\tablescale}{0.98}

\newcommand{\rsub}[1]{\noindent\textbf{#1 ---}\hspace{0.45em}}

\newcommand{\LT}{$\mathcal{L}_T$}
\newcommand{\LP}{$\mathcal{L}_P$}
\newcommand{\LY}{$\mathcal{L}_y$}

\title{Exploring the Design Space of Representation Learning for Audio Transformations}

\multauthor
  {Sungho Lee$^{\natural}$ \hspace{1cm} Marco A. Mart\'inez-Ram\'irez$^\flat$ \hspace{1cm} Junghyun Koo$^\flat$}
  {{\bf Wei-Hsiang Liao$^\flat$ \hspace{1cm} Kyogu Lee$^\natural$ \hspace{1cm} Yuki Mitsufuji$^{\flat\sharp}$}\\
  $^\natural$Department of Intelligence and Information, Seoul National University, Seoul, South Korea\\
  $^\flat$Sony AI, Tokyo, Japan \hspace{0.3cm} $^\sharp$Sony Group Corporation, Tokyo, Japan%
  }

\makeatletter
\if@submission\else
  \def\authorname{S. Lee, M. A. Mart\'inez-Ram\'irez, J. Koo, W.-H. Liao, K. Lee, and Y. Mitsufuji}
\fi
\makeatother

\begin{document}

\maketitle

\begin{abstract}
Neural audio representation learning has enabled a range of content-oriented applications, but the resulting features remain limited for tasks involving audio processing. 
Furthermore, it is not obvious what processing-aware representations should capture: the processing itself, abstracted away from source content, or the processed audio that retains it. 
Existing approaches implicitly commit to one or the other and also differ in their models, data, and evaluation, obscuring which design choices drive their behavior. 
We address both questions within a unified framework of three objectives: processing consistency, description alignment, and equivariance via forward prediction. 
We compare all combinations of the objectives under a controlled setup and reveal their relative strengths and interactions.
Our framework produces both a transformation embedding and a processed-audio embedding, and we find that the two play complementary roles: distance-based tasks favor the former, while probe-based tasks favor the latter. 
Combined with improvements in network architecture and training pipeline, our representations outperform prior baselines across retrieval, probe-based evaluation, and style transfer.
\end{abstract}

\section{Introduction}\label{sec:introduction}

Representation learning for audio has focused on content: models trained on content-oriented objectives~\cite{elizalde2023clap, li2023mert, hershey2017cnn, kong2020panns} either disregard the processing applied to a signal or are made invariant to it through augmentation. Yet processors are what shape sound throughout audio production, from recording and mixing to mastering, and representations that capture them would benefit various downstream tasks such as processor parameter estimation, preset recommendation, and style transfer~\cite{steinmetz2024st, yeh2025fx, koo2023music, moliner2026automatic, combes2025neural, koo2025ito, doh2025llm2fx}.

What such a representation should capture is less obvious than it may appear. A processing-aware representation could mean a representation of the processing itself (e.g., a distortion effect) or the processed audio (e.g., a distorted guitar). These are not the same thing: the former abstracts away source content, while the latter retains it. Existing approaches~\cite{koo2023music, yeh2025fx, steinmetz2024st} implicitly commit to one or the other, but rarely state which, and the consequences of this choice for downstream tasks have not been examined directly.

Prior work also differs in how these representations are learned.
Contrastive methods~\cite{koo2023music, yeh2025fx} shape geometric structure by mapping similarly processed audio close together in the embedding space.
Label-based approaches~\cite{steinmetz2024st} learn representations by predicting annotations of the processing, such as presets.
Equivariance-based methods capture attributes such as pitch and tempo through equivariance to the corresponding transformations~\cite{riou2025pesto, kong2024stone, mccallum2024similar}. The same idea has recently been applied to simple audio processing~\cite{bralios2025re}, but not to complex processing chains.
These approaches further differ in their models, data, and evaluation, which makes direct comparison difficult and confounds the learning objective with the training infrastructure.

In response, we recast these lines of work as three objectives within a single framework: processing consistency via multi-view learning~\cite{chen2020simple}, alignment with parametric processing descriptions~\cite{radford2021learning}, and equivariance via forward prediction~\cite{dangovski2021equivariant, park2022learning}.
Under a controlled setup, we systematically compare all combinations of the three, isolating each objective's contribution to various downstream tasks such as retrieval and probe-based chain and parameter estimation. We find that the three objectives have complementary strengths: processing consistency gives the best invariance to source content, description alignment is stronger overall since the description provides a complete, content-free reference, and equivariance improves the probe-based tasks.

Our framework produces two embeddings: a transformation embedding, which captures the processing as observed in audio, and a processed-audio embedding. The separation between them is necessarily partial, since processing can be observed in audio only through the content-dependent transformations it induces. Nevertheless, such a partial separation is enough to make the two embeddings complementary: distance-based tasks favor the transformation embedding for its processing-oriented geometry, while probe-based tasks favor the processed-audio embedding for its richer information. Finally, combined with improvements in model architecture, batch construction that avoids content shortcuts, and a comprehensive processor library, our framework outperforms prior baselines, with the largest margins under source mismatch.

\section{Framework}\label{sec:method}

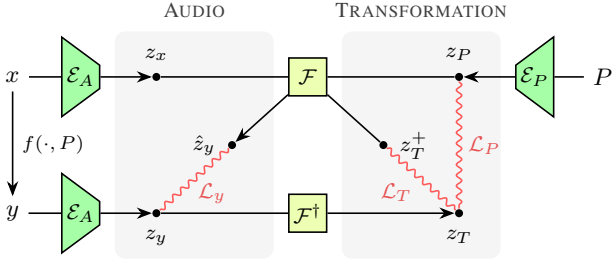
\begin{figure}[!t]
\centering
\setlength{\fboxsep}{0pt}
\input{figures/framework_space_h3}
\vspace{-3.5mm}
\caption{Overview of the proposed framework.}
\label{fig:framework}
\end{figure}

We consider an audio processor $f$ that transforms input audio $x$ into output audio $y = f(x, P)$ according to a description $P$. The processor $f$ is a chain of individual processors, and the description $P$ specifies the type and parameters of each processor in the chain. We distinguish the \emph{processing} (fully specified by $P$) from the content-dependent \emph{transformation} it induces on a given audio signal. A representation derived from audio can therefore capture the processing only through the observed transformation.

There exist two different pretraining settings according to whether the transformation is estimated from the output alone or from both input and output. We refer to them as the \emph{blind}~\cite{koo2023music, yeh2025fx} and \emph{non-blind}~\cite{steinmetz2024st} settings, respectively. Our main experiments use the blind setting; we compare it with non-blind pretraining later.

We explore three objectives that shape the learned representations toward different aspects of the processing; see \figref{fig:framework}. Each objective is a contrastive loss built from a shared formulation~\cite{oord2018representation}: given a query $a$, a positive $b$, and a candidate set $\mathcal{C}$ (including the positive),
\begin{equation}
\ell(a, b; \mathcal{C}) = -\log \frac{\exp(\mathrm{sim}(a, b) / \tau)}{\sum_{c \in \mathcal{C}} \exp(\mathrm{sim}(a, c) / \tau)},
\label{eq:contrastive}
\end{equation}
where $\mathrm{sim}$ denotes cosine similarity and $\tau$ is a learnable temperature. Each loss uses its own $\tau$ and is averaged over all batch items, which serve as both query and candidate.

\medskip
\rsub{Processing Consistency}The first objective encourages embeddings to be similar when the same processing is applied to different sources. An audio encoder $\mathcal{E}_A$ produces the \emph{processed-audio embedding} $z_y = \mathcal{E}_A(y)$. Then, under blind pretraining, an \emph{inverse predictor} $\mathcal{F}^\dagger$ maps $z_y$ to the \emph{transformation embedding} $z_T = \mathcal{F}^\dagger(z_y)$.
Although applying the same processing chain to different sources yields distinct content-dependent transformations, these transformations serve as multiple views of the underlying processing~\cite{chen2020simple}. Accordingly, for each query $z_T$, we derive a positive $z_T^+$ from a different source processed by the same chain, and the processing consistency loss is given by
\begin{equation}
\mathcal{L}_T = \ell(g_T(z_T), g_T(z_T^+); \mathcal{C}_T),
\label{eq:lt}
\end{equation}
where the projection head $g_T$ is applied to the query and positive, and the candidate set $\mathcal{C}_T$ contains all projected transformation embeddings in the batch. This formulation matches Fx-Encoder~\cite{koo2023music} and Fx-Encoder++~\cite{yeh2025fx}, except in how positives and negatives are sampled.

\medskip
\rsub{Description Alignment}
The transformation embedding $z_T$ captures the processing only through the observed transformation; the description $P$ instead specifies the processing in full, providing a complete, content-free reference. To leverage this, a description encoder $\mathcal{E}_P$ maps $P$ to the description embedding $z_P = \mathcal{E}_P(P)$. We then align $z_T$ with $z_P$ using a symmetric contrastive objective~\cite{radford2021learning}:
\begin{equation}
\mathcal{L}_P = \frac{\ell(g_T(z_T), g_P(z_P); \mathcal{C}_P) + \ell(g_P(z_P), g_T(z_T); \mathcal{C}_T)}{2}.
\label{eq:lp}
\end{equation}
The candidate set $\mathcal{C}_P$ contains all projected description embeddings in the batch, while $g_T$ and $\mathcal{C}_T$ are shared with the processing consistency loss.

A related label-based approach, AFx-Rep~\cite{steinmetz2024st}, also associates audio with the underlying processing, but through supervised classification over discrete presets. Our formulation instead employs contrastive alignment with parametric descriptions, extending beyond a fixed vocabulary.

\medskip
\rsub{Equivariance Learning}
The two preceding objectives shape the transformation embedding $z_T$ but do not directly organize the processed-audio embedding $z_y$. We therefore add a third objective defined on $z_y$. Concretely, a \emph{forward predictor} $\mathcal{F}$~\cite{wang2024latent, park2025solving} predicts the processed-audio embedding from the input embedding $z_x = \mathcal{E}_A(x)$ and a conditioning embedding $z_\star$, giving $\hat{z}_{y,\star} = \mathcal{F}(z_x, z_\star)$. The conditioning embedding is either the positive's transformation embedding $z_T^+$ or the description embedding $z_P$; we use $z_T^+$ rather than the query's own $z_T$, because $z_T$ is derived from $z_y$, allowing a trivial identity shortcut. We then apply a symmetric contrastive objective between the prediction $\hat{z}_{y,\star}$ and the observed $z_y$:
\begin{equation}
\begin{split}
\mathcal{L}_y = \sum_{\star \in \mathcal{S}} \frac{1}{2}\big[&\ell(g_y(\hat{z}_{y,\star}), g_y(z_y); \mathcal{C}_y) \\
&{}+ \ell(g_y(z_y), g_y(\hat{z}_{y,\star}); \mathcal{C}_{\hat{y},\star})\big],
\end{split}
\label{eq:ly}
\end{equation}
where the candidate sets $\mathcal{C}_y$ and $\mathcal{C}_{\hat{y},\star}$ contain all projected processed-audio embeddings $g_y(z_y)$ and predicted embeddings $g_y(\hat{z}_{y,\star})$ in the batch, respectively. The projection head $g_y$ is shared across all terms. The conditioning set $\mathcal{S}$ is $\{T^+\}$, extended to $\{T^+, P\}$ when description alignment is used.

Predicting a state from a transformation or action is common in equivariant representation learning~\cite{dangovski2021equivariant, devillers2022equimod, yu2024self} and world models~\cite{park2022learning}. In the audio domain, related work uses closed-form latent actions~\cite{bralios2025re, riou2025pesto, kong2024stone} or learned translations on frozen embeddings~\cite{mccallum2024similar} for specific transformations; since our setting involves chains of nonlinear processors, we use a jointly trained predictor $\mathcal{F}$.

The three objectives thus play distinct roles. The processing consistency loss $\mathcal{L}_T$ shapes processing-oriented geometry by contrasting transformation embeddings with one another, whereas the description alignment loss $\mathcal{L}_P$ injects content-free processing information by pulling $z_T$ toward $z_P$ without such repulsion. The equivariance loss $\mathcal{L}_y$ is the only objective defined on $z_y$, shaping $z_T$ indirectly. All active loss terms nevertheless shape all embeddings, and the encoders and predictors computing them are trained end-to-end.

\section{Implementation}\label{sec:implementation}

\rsub{Model Architecture}
Following~\cite{bralios2025re}, we compose the frozen Stable Audio Open (SAO) encoder~\cite{evans2025stable} with a trainable adapter to form the audio encoder $\mathcal{E}_A$.
Since SAO was trained to reconstruct the full audio signal, its latents retain processing-related information. Further, we can precompute the latents for the entire training set, freeing GPU memory for larger batch sizes.
We reshape the SAO latents to halve the frame rate and double the channel size, then pass them through a 4-layer transformer (1024 hidden dimensions) with rotary positional embeddings~\cite{su2024roformer} and key-value normalization~\cite{dehghani2023scaling}; the mean-pooled output forms the audio embedding ($z_x$ and $z_y \in \mathbb{R}^{1024}$ for $x$ and $y$, respectively).
The inverse predictor $\mathcal{F}^\dagger$ is a 2-layer MLP that produces the transformation embedding $z_T \in \mathbb{R}^{1024}$.

For each processor in the chain, the description encoder $\mathcal{E}_P$ sums a learned type embedding with a parameter embedding. The parameters are normalized to $[0,1]$ and mapped through a type-specific linear layer followed by a shared linear layer for parameter efficiency. The resulting processor embeddings, together with two auxiliary tokens, are passed through a 6-layer transformer (512 hidden dimensions)~\cite{lee2023blind}; the auxiliary token outputs are concatenated to form the description embedding $z_P \in \mathbb{R}^{1024}$.
The forward predictor $\mathcal{F}$ is a 3-layer MLP; all projection heads ($g_T$, $g_P$, $g_y$) are 2-layer MLPs with 128 output dimensions.

\medskip
\rsub{Processor Library}
Because our representations are learned from a joint distribution of source audio and processors, the diversity and quality of the processors are crucial for generalization.
Existing PyTorch audio processing libraries~\cite{steinmetz2024dasp, lee2024grafx, yu2025diffvox, yeh2025difffx} enable GPU-parallel rendering but mostly target differentiable signal processing with limited processor diversity.
In response, we assemble a new library with a broader set of processors, spanning equalizer, filter, dynamics, distortion (including neural amp models~\cite{wright2025open}), reverb (digital filters and impulse-response convolutions), delay, modulation, and more.
Processing chains of length 1--8 are sampled in a hierarchical manner (chain length followed by categories, specific processors, and parameters).
Our implementation and processor library are available at \url{https://github.com/SonyResearch/RLAT}.

\medskip
\rsub{Batch Construction}\label{sec:batch}
Beyond the processor library, the construction of training batches matters because we employ contrastive learning and aim to make the transformation embedding $z_T$ source-invariant.
For example, if the positive shares the ``source identity'' with the query, the model can identify the positive based on content alone.
Despite this, for each query, Fx-Encoder++~\cite{yeh2025fx} uses a same-stem positive drawn from a different segment, and Fx-Encoder~\cite{koo2023music} uses a same-instrument positive.
Conversely, if every negative has a different source identity, the model can distinguish them on source mismatch alone. To prevent both shortcuts, we construct each batch so that every query is assigned a positive that applies the same processing to a different source (possibly from a different instrument). We also include hard negatives that apply different processing to the query's exact source signal.

\section{Evaluation}\label{sec:setup}
\rsub{Setup}
All audio is in stereo at a 44.1 kHz sampling rate.
We train on stems from MedleyDB~\cite{bittner2014medleydb, bittner2016medleydb}, MoisesDB~\cite{pereira2023moisesdb}, and Mixing Secrets~\cite{senior2018mixing}, with all source audio Fx-normalized following prior work~\cite{martinez2022automatic}: the magnitude response, dynamics, and stereo image are matched to the average statistics of each instrument category.
Training processing chains are rendered using our processor library.
For evaluation, we use the test subset of MUSDB~\cite{rafii2017musdb18} as source audio and 67 publicly available Linux audio plugins, loaded via \texttt{Pedalboard}~\cite{sobot_peter_2023_7817838}, with parameters randomized.
We pre-render 1.5M subbatches of 64 tuples (input, output, description) as SAO latents; at training time, 16 subbatches are sampled to form each full batch of 1024. Random crops of 5--10 seconds are drawn from the 16-second latent segments during training.
To systematically ablate the three objectives, we train with the combined loss
\begin{equation}
\mathcal{L} = w_T\,\mathcal{L}_T + w_P\,\mathcal{L}_P + w_y\,\mathcal{L}_y,
\label{eq:total}
\end{equation}
and restrict each weight to $w_T, w_P, w_y \in \{0, 1\}$ so that the seven non-trivial assignments cover all combinations of active terms.
We do not tune the per-loss weights further; increasing the learning rate for single-loss runs to compensate for their smaller gradient magnitude did not improve results. We train for 100k steps using FlashAdamW~\cite{ortiz2026flashoptim} with a peak learning rate of $10^{-4}$ and cosine annealing on a single NVIDIA H100 GPU.

\medskip
\rsub{Episodic Retrieval}We evaluate the geometric structure of the transformation embedding $z_T$: whether similarly processed audio is grouped in the embedding space regardless of content.
We run 20-way episodic retrieval over 1000 episodes per condition, using individual processors (chain length 1) under varying source constraints.
In the \emph{within-stem} (WS) condition, a query, positive, and all negatives share the same stem; in the \emph{within-instrument} (WI) condition, they share the same instrument. This eliminates source variation and isolates the ability to discriminate processing.
In the \emph{cross-stem} (CS) and \emph{cross-instrument} (CI) conditions, the positive uses a different stem or instrument than the query, while all negatives share the query's source; these conditions test source and instrument invariance. An accuracy drop relative to the corresponding within-condition indicates reliance on source or timbral shortcuts rather than processing information.
In the \emph{random} (R) condition, all sources are randomly sampled from the test dataset with no stem/instrument constraints.
\begin{table}[t]
\centering\small
\scalebox{\tablescale}{
\input{tables/retrieval_conditions.tex}
}
\caption{Retrieval accuracy across source conditions.}
\label{tab:retrieval}
\end{table}

\tabref{tab:retrieval} reports retrieval accuracy in percentages alongside results from published baseline checkpoints. Most of our configurations outperform the baselines by a clear margin, particularly under the more challenging cross conditions. Among the baselines, AFx-Rep~\cite{steinmetz2024st} remains competitive on within-stem but falls substantially behind on both cross-stem and cross-instrument conditions.
Both models in the Fx-Encoder family~\cite{yeh2025fx, koo2023music} also struggle under the cross conditions, consistent with their use of same-stem or same-instrument positives that share source identity with the query during training.
General-purpose audio representations~\cite{elizalde2023clap, li2023mert, hershey2017cnn, kong2020panns} are even more vulnerable to content shortcuts, falling below the 5\% chance level on cross-instrument because content-based similarity systematically favors a negative when every negative shares the query's source.
Our configurations also degrade under the cross conditions, but less sharply than the baselines; in the embedding analysis below, we argue that some residual degradation is unavoidable, since the transformation itself depends on content.

\begin{figure}[t]
\centering
\includegraphics[width=\linewidth]{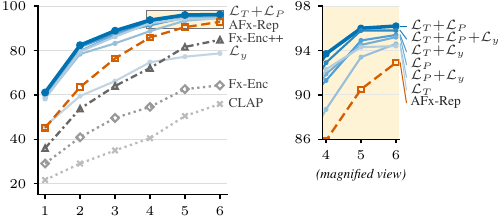}
\caption{Retrieval accuracy across chain lengths.}
\label{fig:chainlen}
\vspace{16pt}
\centering\small
\scalebox{\tablescale}{
\input{tables/generalization_2x2.tex}
}
\par\vspace{5pt}
{\footnotesize Each cell: $\mathcal{L}_T + \mathcal{L}_P$, $\mathcal{L}_T + \mathcal{L}_P + \mathcal{L}_y$, \textcolor{gray}{AFx-Rep~\cite{steinmetz2024st}, and Fx-Enc++~\cite{yeh2025fx}}.\par}
\captionof{table}{Retrieval accuracy along generalization axes.}
\label{tab:gen2x2}
\end{figure}

Next, among our single-loss configurations, description alignment $\mathcal{L}_P$ leads on WS, WI, R, and Avg, consistent with the description embedding $z_P$ being more informative of the processing than the audio-derived $z_T$. Processing consistency $\mathcal{L}_T$, in contrast, is strongest on the harder cross conditions, since only it targets source invariance directly, through positives from a different source and negatives sharing the query's source. Equivariance $\mathcal{L}_y$ is the weakest across all conditions, as it shapes $z_T$ only indirectly.
Combining losses generally improves performance: $\mathcal{L}_T + \mathcal{L}_P$ is the strongest pair overall, and adding $\mathcal{L}_y$ to form the full combination yields a slight drop. The effect of $\mathcal{L}_y$ is asymmetric: it helps $\mathcal{L}_T$ but hurts $\mathcal{L}_P$.

To assess whether the representations scale beyond individual processors, we evaluate them on chains of length up to 6. \figref{fig:chainlen} reports averages across the five retrieval conditions. Accuracy improves with chain length and then saturates quickly, consistent with prior findings~\cite{yeh2025fx}. Most of our configurations outperform the baselines throughout.

We further test generalization along two axes: processor implementation, comparing the seen PyTorch processors against unseen Linux plugins, and source audio distribution, comparing unseen MUSDB sources against sources seen during training. \tabref{tab:gen2x2} reports retrieval results for all four combinations for our two strongest configurations, $\mathcal{L}_T + \mathcal{L}_P$ and $\mathcal{L}_T + \mathcal{L}_P + \mathcal{L}_y$, and two baselines, AFx-Rep and Fx-Encoder++.
The results indicate that the processor gap appears small; the seen PyTorch processors are even slightly harder to distinguish than the unseen Linux plugins, likely because their transformations are subtler. On the other hand, the source gap is much larger, indicating that generalizing to unseen sources is the main bottleneck. Note that the baselines show no such source gap, since they used different training data.

\begin{table*}[t]
\centering\small
\scalebox{\tablescale}{
\input{tables/downstream_merged.tex}
}
\par\vspace{5pt}
{\footnotesize EQ: equalizer; Filt.: filter; Dyn.: dynamics; Dist.: distortion; Rev.: reverb; Del.: delay; and Mod.: modulation.\par}
\caption{Downstream task and source classification results with frozen embeddings and MLP probes.}
\label{tab:downstream}
\end{table*}

\rsub{Probe Evaluation}In practice, only the processed audio is typically available, making blind estimation of processing attributes a natural downstream task. We therefore evaluate whether a probe can predict these attributes from frozen embeddings.
Unlike retrieval, which relies on the processing-oriented geometry of the transformation embedding $z_T$, we probe the processed-audio embedding $z_y$: it contains both source and processing information, and the probe must learn to extract the processing-relevant component. We justify this choice in the following embedding analysis.
We train 3-layer MLPs on processed-audio embeddings from MUSDB (3600 training, 400 validation, and 1000 test samples of 10 seconds from non-overlapping songs; batch size 64) until the validation loss converges.

We first evaluate chain estimation, in which the probe predicts which processor categories were applied (chain lengths from 1 to 6) using multi-label classification. We report intersection-over-union (IoU), exact-match accuracy, and F1 score (macro and per-category). \tabref{tab:downstream} shows the results. The full loss combination achieves the best results across all three metrics. The equivariance loss $\mathcal{L}_y$, which showed limited benefit for retrieval, proves valuable here through its direct effect on $z_y$. 

We next evaluate parameter estimation: the probe predicts normalized processor parameters for representative plugins (one per category), and we report the mean absolute error (MAE, $\times$100). Most configurations match or outperform the baselines, though margins are smaller than for chain estimation.

\begin{figure}[t]
\centering
\includegraphics[width=.93\columnwidth]{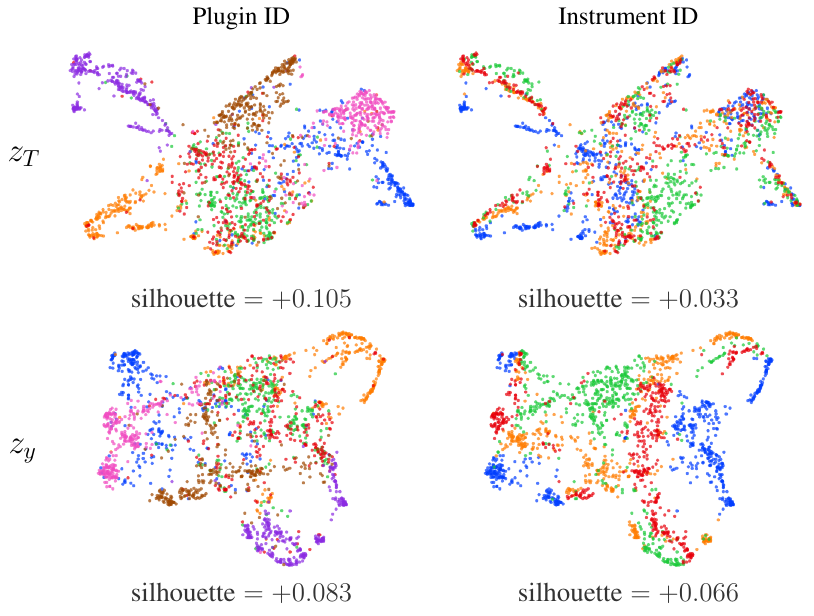}
\caption{UMAP visualizations of embeddings.}
\label{fig:umaps}
\end{figure}

\medskip
\rsub{Embedding Analysis}We now compare the two embeddings directly, in source content, geometry, and downstream behavior. The transformation embedding $z_T$ is designed to encode the processing rather than the source content, and the processed-audio embedding $z_y$ to retain both. To measure this, we train a 4-way stem classifier (drums, bass, vocals, and other) on each, under the same setup as the probe evaluation; lower accuracy indicates less source content.

The two rightmost columns of \tabref{tab:downstream} report the results. For single-embedding baselines, we report only the relevant column. The Fx-Encoder family~\cite{koo2023music, yeh2025fx} uses blind contrastive training, same as our processing consistency setup, and produces an embedding that corresponds to $z_T$. AFx-Rep~\cite{steinmetz2024st} and the general-purpose baselines produce a single embedding of the processed audio, which corresponds to $z_y$.
The processing consistency configuration achieves the lowest source leakage for $z_T$, below the Fx-Encoder family despite sharing the same contrastive loss, attributable to our hard-negative batch construction. Adding the equivariance loss $\mathcal{L}_y$ increases leakage, as forward prediction encourages the preservation of source information.
Most audio processors are nonlinear and thus signal-dependent: a compressor applied to drums induces a different transformation than the same compressor on vocals. Therefore, even though the contrastive objective with hard negatives reduces content shortcuts, it cannot eliminate the content the transformation inherently depends on, so some source leakage in $z_T$ is expected.

\figref{fig:umaps} shows UMAP~\cite{mcinnes2018umap} visualizations of $z_T$ and $z_y$ for the full loss configuration, colored by plugin (left) and source instrument (right), with silhouette scores~\cite{rousseeuw1987silhouettes} computed in the original embedding space. Instrument-related grouping is clearer in $z_y$, whereas plugin-related grouping is clearer in $z_T$. Yet the stem classifier above recovers source identity from $z_T$, showing that geometric organization and probe-based recoverability are different properties.

\begin{table}[t]
\centering\small
\scalebox{\tablescale}{
\input{tables/embedding_comparison.tex}
}
\caption{Embedding and pretraining comparison.}
\label{tab:emb_comparison}
\end{table}

Finally, we compare the two embeddings on the downstream tasks; see the first two columns of \tabref{tab:emb_comparison}. Since $z_T$ is derived from $z_y$, the data processing inequality~\cite{cover1991elements} implies that it cannot contain more information about the source or processing than $z_y$. Accordingly, $z_y$ outperforms $z_T$ on all probe-based metrics as a nonlinear probe can extract processing from the richer representation, justifying our choice of $z_y$ in the probe evaluation. For episodic retrieval, $z_y$ matches $z_T$ on within-stem but struggles under the cross conditions. Since $z_y$ is not shaped toward source invariance, the content it retains favors the negatives.

\medskip
\rsub{Non-blind Pretraining}\label{sec:blind}
Our main experiments use the blind setting, matching the Fx-Encoder family~\cite{koo2023music, yeh2025fx}.
Here, we additionally compare with non-blind pretraining, which AFx-Rep~\cite{steinmetz2024st} uses.
In the non-blind setting, the inverse predictor $\mathcal{F}^\dagger$ takes the input audio embedding $z_x$ alongside the processed-audio embedding $z_y$, concatenated as input.
The distinction concerns pretraining only: inference is blind in both cases, as only the processed audio is available.
The non-blind model must therefore predict $z_x$ at test time.
To this end, a 2-layer transformer estimates it from the frame-level features of the processed audio before mean pooling, trained on detached features so that it does not shape the representation.
$\mathcal{F}^\dagger$ then produces the estimated transformation embedding $\hat{z}_T$.

\tabref{tab:emb_comparison} reports results on the four representations available at inference: $z_T$ and $z_y$ under blind pretraining, $\hat{z}_T$ and $z_y$ under non-blind.
We additionally report a non-blind oracle $z_T$, computed with the ground-truth $z_x$, as an upper bound.
Of these, AFx-Rep conceptually corresponds to the non-blind $z_y$.
Because the input is given during pretraining, $z_y$ is under no pressure to encode the processing invariantly to content, and it accordingly falls well below its blind counterpart on the cross conditions.
Blind $z_T$ retains a small advantage over $\hat{z}_T$, suggesting end-to-end training under the blind constraint is more effective than a separate input-estimation head.
The upper bound shows substantial headroom across all conditions, indicating input estimation as a key bottleneck for further improvement.

\begin{table}[t]
\centering\small
\scalebox{\tablescale}{
\input{tables/stito_results.tex}
}
\caption{Style transfer results.}
\label{tab:stito}
\end{table}

\medskip
\rsub{Style Transfer}\label{sec:style_transfer}
Finally, we apply the representations to audio style transfer via inference-time parameter optimization~\cite{steinmetz2024st}.
Given source audio and target audio processed by a processor, a black-box optimizer~\cite{hansen1996adapting} searches for parameters that minimize the cosine distance between the candidate's and target's embeddings, using either the transformation embedding $z_T$ or the processed-audio embedding $z_y$ of the full loss configuration $\mathcal{L}_T + \mathcal{L}_P + \mathcal{L}_y$.
We test across three conditions similar to retrieval but with source and target replacing query and positive: within-stem (WS), within-instrument (WI), and cross-instrument (CI).
To isolate the effect of source pairing, each test case shares the same processor across all three conditions.
We test on 1000 source-target pairs (5 seconds) using the evaluation plugins and report the stereo multi-resolution STFT distance (MRSTFT)~\cite{yamamoto2020parallel, lee2025reverse}.
Lower MRSTFT indicates that embedding distance serves as a better optimization surrogate for audio similarity.

\tabref{tab:stito} shows the average MRSTFT results with 95\% confidence intervals.
On within-stem, $z_T$, $z_y$, and AFx-Rep~\cite{steinmetz2024st} are comparable (paired $t$-test, $p > 0.2$): the shared source eliminates the need for invariance, so a processing-oriented geometry offers no advantage.
As source mismatch increases, $z_T$ significantly outperforms both $z_y$ and AFx-Rep by a widening margin ($p < 0.001$), confirming that its geometry provides a functional advantage for cross-source style transfer.

\section{Discussion}\label{sec:discussion}

The transformation embedding $z_T$ and the processed-audio embedding $z_y$ are each designed to capture their respective targets, but complete separation is not achievable when the transformation is content-dependent. We therefore observe a partial separation: both embeddings encode processing similarity, but it is the primary organizing principle only in $z_T$, while $z_y$ also retains source information. Since $z_T$ is derived from $z_y$, its advantage is not information content but geometric organization. This explains the consistent pattern across evaluations: distance-based tasks (retrieval, cross-source style transfer) prefer $z_T$, while probe-based tasks prefer $z_y$. The style transfer results further show that this geometry is not only useful for retrieval but also affects optimization in a downstream audio processing task.

Within this two-embedding picture, we compared all combinations of the three objectives under a fixed experimental setup. Among these, the processing consistency and description alignment pair $\mathcal{L}_T + \mathcal{L}_P$ was strongest for retrieval, while the full combination $\mathcal{L}_T + \mathcal{L}_P + \mathcal{L}_y$ performed best on the probe-based tasks. The objectives also interacted asymmetrically: adding equivariance $\mathcal{L}_y$ to processing consistency $\mathcal{L}_T$ improved performance, whereas adding it to description alignment $\mathcal{L}_P$ degraded results slightly. When paired with $\mathcal{L}_T$, $\mathcal{L}_P$ contributed more than $\mathcal{L}_y$, which contrasts with findings in vision~\cite{yu2024self}: audio processors are nonlinear and chained, so forward prediction is a substantially harder learning signal than the simple augmentations used in vision. Taken together, the ablations clarify the objectives' distinct roles: $\mathcal{L}_T$ directly organizes $z_T$ and improves source robustness, $\mathcal{L}_P$ aligns $z_T$ with the complete, content-free description embedding $z_P$, and $\mathcal{L}_y$ directly shapes $z_y$, benefiting probe-based tasks but indirectly increasing source leakage in $z_T$.

Beyond the objective comparison, several practical concerns remain. The retrieval results indicate that generalization is limited by the diversity of the source audio. Fx-normalization is a domain-specific preprocessing step that requires source categories, limiting its scalability beyond controlled settings. Since the non-blind setup observes the input audio directly, it requires no assumptions about the characteristics of the unprocessed input and could enable pretraining with larger-scale raw data. Even then, recovering $z_T$ from the processed audio alone remains a blind estimation problem, which could be treated as a separate downstream task; Fx-normalization would be needed only there. A separate concern is $\mathcal{L}_P$, which requires structured processor descriptions that are unavailable for some real-world plugins; the processing consistency and equivariance pair $\mathcal{L}_T + \mathcal{L}_y$ then serves as a practical fallback.

A complete representation of audio processing involves three aspects: the source content, the transformation, and the transformed audio. We addressed the latter two with distinct embeddings, and demonstrated that such embeddings should be not only informative but also geometrically organized for the downstream task. Extending the framework to capture all three aspects in a single representation model remains an open direction.

\newpage

\section{Acknowledgments}
This work was done during the first author's internship at Sony AI. This work was also supported by the InnoCORE program of the Ministry of Science and ICT (AI Meta-Scientist, N10260110).
\bibliography{ISMIRtemplate}

\ifpreprint
\appendix
\section{Per-Objective Views of the Framework}
\label{sec:appendix-objviews}
Each panel highlights the components active for one objective; the rest of the framework (\figref{fig:framework}) is faded.
\begin{figure}[h]
\centering
\begin{subfigure}{\columnwidth}\centering\input{figures/framework_LT}\caption{Processing consistency $\mathcal{L}_T$.}\end{subfigure}\\[6pt]
\begin{subfigure}{\columnwidth}\centering\input{figures/framework_LP}\caption{Description alignment $\mathcal{L}_P$.}\end{subfigure}\\[6pt]
\begin{subfigure}{\columnwidth}\centering\input{figures/framework_LyT}\caption{Equivariance $\mathcal{L}_y$, conditioned on $z_T^+$.}\end{subfigure}\\[6pt]
\begin{subfigure}{\columnwidth}\centering\input{figures/framework_LyP}\caption{Equivariance $\mathcal{L}_y$, conditioned on $z_P$.}\end{subfigure}
\caption{Per-objective views of the framework.}
\label{fig:appendix-objviews}
\end{figure}
\fi

\end{document}

%% file: figures/framework_space_h3.tex
\colorlet{encodercolor}{green!35}
\colorlet{predictorcolor}{lime!30}
\colorlet{losscolor}{red!65}
\colorlet{spacecolor}{gray!8}

\begin{tikzpicture}[scale=1.0, every node/.style={scale=1.0},
    >=Stealth,
    font=\footnotesize,
    dot/.style={circle, fill=black, minimum size=3pt, inner sep=0pt},
    encoder/.style={draw, semithick, fill=encodercolor, trapezium,
        shape border rotate=90,
        trapezium left angle=130, trapezium right angle=130,
        trapezium stretches body,
        minimum width=9mm, minimum height=5mm, inner sep=1pt},
    dec/.style={draw, semithick, fill=predictorcolor, minimum size=5mm, inner sep=0.5pt},
    arrow/.style={->, semithick},
    lossarrow/.style={decorate, decoration={snake, amplitude=1pt, segment length=4pt}, semithick, losscolor},
    losslabel/.style={font=\footnotesize, text=losscolor!90!black},
]

\def\dx{0.1}
\def\encX{1.0}
\def\ax{2.5}
\def\axBot{2}
\def\axTop{3}
\def\midX{4}
\def\tx{5.5}
\def\txBot{5}
\def\txTop{6}
\def\ePX{7}
\def\pX{7.9}

\def\ly{-0.6}
\def\my{0.3}
\def\ry{1.2}

\def\spaceW{2.1cm}
\def\spaceH{3cm}

\node[font=\footnotesize\scshape, text=gray!50!black, anchor=south] at (\ax, \my + 1.56) {Audio};
\node[font=\footnotesize\scshape, text=gray!50!black, anchor=south] at (\tx, \my + 1.56) {Transformation};

\node[dot] (zTp) at (\txBot, \my) {};
\node (zTpLbl) [right=1pt of zTp] {$z_T^+$};

\node[dot] (zT) at (\txTop, \ly) {};
\node (zTLbl) [below=1pt of zT] {$z_T$};

\node[dot] (zx) at (\axBot, \ry) {};
\node (zxLbl) [above=1pt of zx] {$z_x$};

\node[dot] (zy) at (\axBot, \ly) {};
\node (zyLbl) [below=1pt of zy] {$z_y$};

\node[dot] (zyhat) at (\axTop, \my) {};
\node (zyhatLbl) [left=1pt of zyhat] {$\hat{z}_y$};

\node[dot] (zP) at (\txTop, \ry) {};
\node (zPLbl) [above=1pt of zP] {$z_P$};

\begin{scope}[on background layer]
    \node[fill=spacecolor, rounded corners=4pt, minimum width=\spaceW, minimum height=\spaceH] at (\ax, \my) {};
    \node[fill=spacecolor, rounded corners=4pt, minimum width=\spaceW, minimum height=\spaceH] at (\tx, \my) {};
\end{scope}

\draw[lossarrow] (zP) -- (zT);
\node[losslabel, right=0pt] at ($(zP)!0.5!(zT)$) {$\mathcal{L}_P$};

\draw[lossarrow] (zTp) -- (zT);
\node[losslabel, below left=-2pt and 0pt] at ($(zTp)!0.5!(zT)$) {$\mathcal{L}_T$};

\draw[lossarrow] (zyhat) -- (zy);
\node[losslabel, below right=-2pt and -2pt] at ($(zyhat)!0.5!(zy)$) {$\mathcal{L}_y$};

\node[dec] (Fdagger) at (\midX, \ly) {$\mathcal{F}^\dagger$};
\draw[semithick] (zy) -- (Fdagger);
\draw[arrow] (Fdagger) -- (zT);

\node[dec] (Fbox) at (\midX, \ry) {$\mathcal{F}$};
\draw[semithick] (zx) -- (Fbox);
\draw[arrow] (Fbox) -- (zyhat);
\draw[semithick] (zTp) -- (Fbox);
\draw[semithick] (zP) -- (Fbox);

\node[font=\small] (x) at (\dx, \ry) {$x$};
\node[font=\small] (y) at (\dx, \ly) {$y$};
\draw[arrow] (x) -- node[right, font=\scriptsize] {$f(\cdot, P)$} (y);

\node[encoder] (eaX) at (\encX, \ry) {$\mathcal{E}_A$};
\draw[semithick] (x) -- (eaX);
\draw[arrow] (eaX) -- (zx);

\node[encoder] (eaY) at (\encX, \ly) {$\mathcal{E}_A$};
\draw[semithick] (y) -- (eaY);
\draw[arrow] (eaY) -- (zy);

\node[font=\small] (P) at (\pX, \ry) {$P$};
\node[encoder, shape border rotate=270] (eP) at (\ePX, \ry) {$\mathcal{E}_P$};
\draw[semithick] (P) -- (eP);
\draw[arrow] (eP) -- (zP);

\end{tikzpicture}

%% file: tables/retrieval_conditions.tex
\setlength{\tabcolsep}{4.5pt}%
\begin{tabular}{l c c c c c c}
\toprule
 & \multicolumn{6}{c}{Condition} \\
\cmidrule(lr){2-7}
Configuration & WS & WI & CS & CI & R & Avg \\
\midrule
\LT & 83.8 & 57.6 & \textbf{46.3} & 48.8 & 54.4 & 58.2 \\
\LP & \underline{87.7} & 62.6 & 43.5 & 46.6 & \underline{58.1} & 59.7 \\
\LY & 81.8 & 50.7 & 26.4 & 24.4 & 46.0 & 45.9 \\
\LT\hspace{3pt}$+$\hspace{3pt}\LP & 87.1 & \textbf{64.3} & 45.4 & \textbf{49.8} & \textbf{58.7} & \textbf{61.1} \\
\LT\hspace{3pt}$+$\hspace{3pt}\LY & 86.6 & 61.5 & \underline{46.0} & \underline{49.2} & 57.8 & 60.2 \\
\LP\hspace{3pt}$+$\hspace{3pt}\LY & \textbf{89.2} & \underline{63.9} & 42.4 & 40.7 & 57.5 & 58.7 \\
\LT\hspace{3pt}$+$\hspace{3pt}\LP\hspace{3pt}$+$\hspace{3pt}\LY & 87.3 & 63.2 & 45.5 & 48.7 & 57.3 & \underline{60.4} \\
\arrayrulecolor{darkgray}\midrule\arrayrulecolor{black}
AFx-Rep \cite{steinmetz2024st} & \underline{87.7} & 51.4 & 28.4 & 20.5 & 38.1 & 45.2 \\
Fx-Encoder++ \cite{yeh2025fx} & 84.9 & 36.9 & 19.9 & 13.2 & 25.0 & 36.0 \\
Fx-Encoder \cite{koo2023music} & 67.0 & 30.6 & 23.8 & \phantom{0}6.7 & 17.6 & 29.1 \\
\arrayrulecolor{darkgray}\midrule\arrayrulecolor{black}
CLAP \cite{elizalde2023clap} & 68.6 & 19.1 & \phantom{0}7.5 & \phantom{0}1.0 & 12.1 & 21.7 \\
MERT \cite{li2023mert} & 65.1 & 17.0 & \phantom{0}5.2 & \phantom{0}2.6 & 11.8 & 20.3 \\
VGGish \cite{hershey2017cnn} & 63.5 & 15.6 & \phantom{0}9.0 & \phantom{0}1.6 & 10.6 & 20.1 \\
PANN \cite{kong2020panns} & 56.9 & 12.4 & \phantom{0}5.8 & \phantom{0}1.7 & \phantom{0}9.7 & 17.3 \\
\bottomrule
\end{tabular}

%% file: tables/generalization_2x2.tex
\begin{tabular}{lcc}
\toprule
 & \multicolumn{2}{c}{Source audio} \\
\cmidrule(lr){2-3}
Processors & Unseen & Seen \\
\midrule
Unseen & \textbf{61.1}~~\underline{60.4}~~\textcolor{gray}{45.2~~36.0} & \textbf{83.2}~~\underline{81.7}~~\textcolor{gray}{42.8~~36.6} \\
Seen & \textbf{58.1}~~\underline{56.3}~~\textcolor{gray}{34.6~~30.6} & \underline{78.4}~~\textbf{78.5}~~\textcolor{gray}{35.7~~30.5} \\
\bottomrule
\end{tabular}

%% file: tables/downstream_merged.tex
\begin{tabular}{lccc@{\hskip 2.9pt}c@{\hskip 2.9pt}c@{\hskip 2.9pt}c@{\hskip 2.9pt}c@{\hskip 2.9pt}c@{\hskip 2.9pt}c@{\hskip 2.9pt}c c@{\hskip 2.9pt}c@{\hskip 2.9pt}c@{\hskip 2.9pt}c@{\hskip 2.9pt}c@{\hskip 2.9pt}c@{\hskip 2.9pt}c@{\hskip 2.9pt}c cc}
\toprule
 & \multicolumn{10}{c}{Chain estimation} & \multicolumn{8}{c}{Parameter estimation} & \multicolumn{2}{c}{Source cls.} \\
\cmidrule(lr){2-11} \cmidrule(lr){12-19} \cmidrule(lr){20-21}
Configuration & IoU & Exact & F1 & {\scriptsize\color{darkgray}EQ} & {\scriptsize\color{darkgray}Filt.} & {\scriptsize\color{darkgray}Dyn.} & {\scriptsize\color{darkgray}Dist.} & {\scriptsize\color{darkgray}Rev.} & {\scriptsize\color{darkgray}Del.} & {\scriptsize\color{darkgray}Mod.} & MAE & {\scriptsize\color{darkgray}EQ} & {\scriptsize\color{darkgray}Filt.} & {\scriptsize\color{darkgray}Dyn.} & {\scriptsize\color{darkgray}Dist.} & {\scriptsize\color{darkgray}Rev.} & {\scriptsize\color{darkgray}Del.} & {\scriptsize\color{darkgray}Mod.} & $z_T$ & $z_y$ \\
\midrule
\LT & 51.5 & 16.1 & 58.3 & {\scriptsize\color{darkgray}45.4} & {\scriptsize\color{darkgray}76.1} & {\scriptsize\color{darkgray}52.1} & {\scriptsize\color{darkgray}30.8} & {\scriptsize\color{darkgray}\textbf{56.2}} & {\scriptsize\color{darkgray}74.0} & {\scriptsize\color{darkgray}73.6} & 17.7 & {\scriptsize\color{darkgray}\textbf{21.5}} & {\scriptsize\color{darkgray}21.0} & {\scriptsize\color{darkgray}18.7} & {\scriptsize\color{darkgray}\phantom{0}\textbf{5.1}} & {\scriptsize\color{darkgray}22.5} & {\scriptsize\color{darkgray}19.6} & {\scriptsize\color{darkgray}15.2} & \textbf{67.2} & 93.7 \\
\LP & \underline{53.3} & \underline{17.8} & 59.6 & {\scriptsize\color{darkgray}47.9} & {\scriptsize\color{darkgray}\textbf{78.0}} & {\scriptsize\color{darkgray}55.7} & {\scriptsize\color{darkgray}29.0} & {\scriptsize\color{darkgray}55.5} & {\scriptsize\color{darkgray}74.5} & {\scriptsize\color{darkgray}76.3} & \textbf{17.5} & {\scriptsize\color{darkgray}\textbf{21.5}} & {\scriptsize\color{darkgray}\textbf{20.8}} & {\scriptsize\color{darkgray}\textbf{18.5}} & {\scriptsize\color{darkgray}\phantom{0}5.3} & {\scriptsize\color{darkgray}22.0} & {\scriptsize\color{darkgray}\textbf{19.2}} & {\scriptsize\color{darkgray}\underline{14.9}} & \underline{83.1} & 93.2 \\
\LY & 50.5 & 15.4 & 57.6 & {\scriptsize\color{darkgray}46.7} & {\scriptsize\color{darkgray}76.0} & {\scriptsize\color{darkgray}50.7} & {\scriptsize\color{darkgray}29.1} & {\scriptsize\color{darkgray}52.5} & {\scriptsize\color{darkgray}73.2} & {\scriptsize\color{darkgray}75.4} & 17.7 & {\scriptsize\color{darkgray}\underline{21.7}} & {\scriptsize\color{darkgray}\underline{20.9}} & {\scriptsize\color{darkgray}18.8} & {\scriptsize\color{darkgray}\phantom{0}5.4} & {\scriptsize\color{darkgray}22.3} & {\scriptsize\color{darkgray}19.8} & {\scriptsize\color{darkgray}15.0} & 93.5 & \underline{94.9} \\
\LT\hspace{3pt}$+$\hspace{3pt}\LP & 53.0 & 17.3 & \underline{59.9} & {\scriptsize\color{darkgray}\underline{49.1}} & {\scriptsize\color{darkgray}\underline{77.3}} & {\scriptsize\color{darkgray}55.6} & {\scriptsize\color{darkgray}30.4} & {\scriptsize\color{darkgray}55.3} & {\scriptsize\color{darkgray}\textbf{74.8}} & {\scriptsize\color{darkgray}76.5} & \textbf{17.5} & {\scriptsize\color{darkgray}\textbf{21.5}} & {\scriptsize\color{darkgray}\textbf{20.8}} & {\scriptsize\color{darkgray}\underline{18.6}} & {\scriptsize\color{darkgray}\phantom{0}5.3} & {\scriptsize\color{darkgray}22.0} & {\scriptsize\color{darkgray}\textbf{19.2}} & {\scriptsize\color{darkgray}15.0} & 85.0 & 94.0 \\
\LT\hspace{3pt}$+$\hspace{3pt}\LY & 52.7 & 16.5 & 59.6 & {\scriptsize\color{darkgray}44.4} & {\scriptsize\color{darkgray}76.9} & {\scriptsize\color{darkgray}\textbf{57.7}} & {\scriptsize\color{darkgray}\underline{32.7}} & {\scriptsize\color{darkgray}55.1} & {\scriptsize\color{darkgray}\underline{74.6}} & {\scriptsize\color{darkgray}75.9} & \underline{17.6} & {\scriptsize\color{darkgray}\textbf{21.5}} & {\scriptsize\color{darkgray}21.0} & {\scriptsize\color{darkgray}18.7} & {\scriptsize\color{darkgray}\phantom{0}\underline{5.2}} & {\scriptsize\color{darkgray}22.3} & {\scriptsize\color{darkgray}\underline{19.4}} & {\scriptsize\color{darkgray}15.0} & 88.9 & 94.0 \\
\LP\hspace{3pt}$+$\hspace{3pt}\LY & 52.9 & 17.5 & 59.3 & {\scriptsize\color{darkgray}48.5} & {\scriptsize\color{darkgray}77.2} & {\scriptsize\color{darkgray}54.5} & {\scriptsize\color{darkgray}28.6} & {\scriptsize\color{darkgray}54.3} & {\scriptsize\color{darkgray}74.2} & {\scriptsize\color{darkgray}\underline{77.9}} & \textbf{17.5} & {\scriptsize\color{darkgray}\textbf{21.5}} & {\scriptsize\color{darkgray}21.0} & {\scriptsize\color{darkgray}\underline{18.6}} & {\scriptsize\color{darkgray}\phantom{0}\underline{5.2}} & {\scriptsize\color{darkgray}\underline{21.9}} & {\scriptsize\color{darkgray}\underline{19.4}} & {\scriptsize\color{darkgray}\underline{14.9}} & 90.6 & 94.6 \\
\LT\hspace{3pt}$+$\hspace{3pt}\LP\hspace{3pt}$+$\hspace{3pt}\LY & \textbf{53.8} & \textbf{18.1} & \textbf{60.9} & {\scriptsize\color{darkgray}\textbf{51.6}} & {\scriptsize\color{darkgray}77.2} & {\scriptsize\color{darkgray}\underline{57.3}} & {\scriptsize\color{darkgray}\textbf{32.8}} & {\scriptsize\color{darkgray}\underline{55.9}} & {\scriptsize\color{darkgray}\textbf{74.8}} & {\scriptsize\color{darkgray}76.6} & \textbf{17.5} & {\scriptsize\color{darkgray}\textbf{21.5}} & {\scriptsize\color{darkgray}\underline{20.9}} & {\scriptsize\color{darkgray}\textbf{18.5}} & {\scriptsize\color{darkgray}\phantom{0}\underline{5.2}} & {\scriptsize\color{darkgray}22.0} & {\scriptsize\color{darkgray}\underline{19.4}} & {\scriptsize\color{darkgray}15.0} & 89.4 & \textbf{95.1} \\
\arrayrulecolor{darkgray}\midrule\arrayrulecolor{black}
AFx-Rep \cite{steinmetz2024st} & 50.9 & 14.6 & 56.1 & {\scriptsize\color{darkgray}24.6} & {\scriptsize\color{darkgray}76.5} & {\scriptsize\color{darkgray}55.7} & {\scriptsize\color{darkgray}30.9} & {\scriptsize\color{darkgray}53.5} & {\scriptsize\color{darkgray}73.6} & {\scriptsize\color{darkgray}\textbf{78.1}} & 17.7 & {\scriptsize\color{darkgray}22.0} & {\scriptsize\color{darkgray}21.1} & {\scriptsize\color{darkgray}18.8} & {\scriptsize\color{darkgray}\phantom{0}5.4} & {\scriptsize\color{darkgray}\textbf{21.2}} & {\scriptsize\color{darkgray}20.8} & {\scriptsize\color{darkgray}\textbf{14.6}} & $-$ & 92.2 \\
Fx-Encoder++ \cite{yeh2025fx} & 41.6 & \phantom{0}9.0 & 48.1 & {\scriptsize\color{darkgray}20.0} & {\scriptsize\color{darkgray}68.8} & {\scriptsize\color{darkgray}35.4} & {\scriptsize\color{darkgray}26.4} & {\scriptsize\color{darkgray}45.9} & {\scriptsize\color{darkgray}68.1} & {\scriptsize\color{darkgray}71.8} & 18.8 & {\scriptsize\color{darkgray}22.5} & {\scriptsize\color{darkgray}21.4} & {\scriptsize\color{darkgray}19.4} & {\scriptsize\color{darkgray}\phantom{0}5.8} & {\scriptsize\color{darkgray}23.8} & {\scriptsize\color{darkgray}22.3} & {\scriptsize\color{darkgray}16.3} & 89.4 & $-$ \\
Fx-Encoder \cite{koo2023music} & 36.6 & \phantom{0}6.4 & 40.5 & {\scriptsize\color{darkgray}22.2} & {\scriptsize\color{darkgray}65.7} & {\scriptsize\color{darkgray}43.9} & {\scriptsize\color{darkgray}\phantom{0}1.2} & {\scriptsize\color{darkgray}24.8} & {\scriptsize\color{darkgray}61.3} & {\scriptsize\color{darkgray}64.3} & 19.1 & {\scriptsize\color{darkgray}22.3} & {\scriptsize\color{darkgray}22.7} & {\scriptsize\color{darkgray}19.2} & {\scriptsize\color{darkgray}\phantom{0}6.1} & {\scriptsize\color{darkgray}24.0} & {\scriptsize\color{darkgray}22.2} & {\scriptsize\color{darkgray}17.2} & 89.2 & $-$ \\
\arrayrulecolor{darkgray}\midrule\arrayrulecolor{black}
CLAP \cite{elizalde2023clap} & 38.9 & \phantom{0}9.1 & 39.1 & {\scriptsize\color{darkgray}\phantom{0}7.2} & {\scriptsize\color{darkgray}71.7} & {\scriptsize\color{darkgray}12.9} & {\scriptsize\color{darkgray}\phantom{0}0.0} & {\scriptsize\color{darkgray}37.1} & {\scriptsize\color{darkgray}69.8} & {\scriptsize\color{darkgray}74.9} & 19.2 & {\scriptsize\color{darkgray}22.6} & {\scriptsize\color{darkgray}21.6} & {\scriptsize\color{darkgray}19.4} & {\scriptsize\color{darkgray}\phantom{0}5.9} & {\scriptsize\color{darkgray}23.6} & {\scriptsize\color{darkgray}23.0} & {\scriptsize\color{darkgray}18.3} & $-$ & \underline{94.9} \\
MERT \cite{li2023mert} & 41.6 & \phantom{0}8.5 & 42.8 & {\scriptsize\color{darkgray}\phantom{0}0.3} & {\scriptsize\color{darkgray}71.4} & {\scriptsize\color{darkgray}33.3} & {\scriptsize\color{darkgray}26.7} & {\scriptsize\color{darkgray}23.0} & {\scriptsize\color{darkgray}71.7} & {\scriptsize\color{darkgray}73.3} & 19.4 & {\scriptsize\color{darkgray}22.6} & {\scriptsize\color{darkgray}21.6} & {\scriptsize\color{darkgray}19.4} & {\scriptsize\color{darkgray}\phantom{0}5.6} & {\scriptsize\color{darkgray}23.9} & {\scriptsize\color{darkgray}23.5} & {\scriptsize\color{darkgray}18.9} & $-$ & 91.5 \\
VGGish \cite{hershey2017cnn} & 32.7 & \phantom{0}7.0 & 30.2 & {\scriptsize\color{darkgray}\phantom{0}0.0} & {\scriptsize\color{darkgray}68.2} & {\scriptsize\color{darkgray}\phantom{0}1.1} & {\scriptsize\color{darkgray}\phantom{0}0.0} & {\scriptsize\color{darkgray}\phantom{0}6.5} & {\scriptsize\color{darkgray}65.0} & {\scriptsize\color{darkgray}70.9} & 19.7 & {\scriptsize\color{darkgray}22.7} & {\scriptsize\color{darkgray}22.1} & {\scriptsize\color{darkgray}19.6} & {\scriptsize\color{darkgray}\phantom{0}6.4} & {\scriptsize\color{darkgray}24.0} & {\scriptsize\color{darkgray}23.8} & {\scriptsize\color{darkgray}19.6} & $-$ & 93.7 \\
PANN \cite{kong2020panns} & 37.1 & \phantom{0}7.7 & 35.1 & {\scriptsize\color{darkgray}\phantom{0}0.2} & {\scriptsize\color{darkgray}69.6} & {\scriptsize\color{darkgray}18.2} & {\scriptsize\color{darkgray}\phantom{0}0.6} & {\scriptsize\color{darkgray}14.7} & {\scriptsize\color{darkgray}70.6} & {\scriptsize\color{darkgray}72.0} & 19.7 & {\scriptsize\color{darkgray}22.7} & {\scriptsize\color{darkgray}22.1} & {\scriptsize\color{darkgray}19.6} & {\scriptsize\color{darkgray}\phantom{0}6.4} & {\scriptsize\color{darkgray}24.1} & {\scriptsize\color{darkgray}24.0} & {\scriptsize\color{darkgray}19.4} & $-$ & 93.8 \\
\bottomrule
\end{tabular}

%% file: tables/embedding_comparison.tex
\begin{tabular}{lcc cc c}
\toprule
 & \multicolumn{2}{c}{Blind} & \multicolumn{3}{c}{Non-blind} \\
\cmidrule(lr){2-3} \cmidrule(lr){4-6}
Metric & $z_T$ & $z_y$ & $\hat{z}_{T}$ & $z_y$ & $z_T$ \\
\midrule
Accuracy (WS) & 87.3 & \underline{87.7} & \textbf{88.2} & 86.3 & 98.0 \\
Accuracy (WI) & \textbf{63.2} & \underline{62.5} & 62.3 & 55.4 & 94.7 \\
Accuracy (CS) & \textbf{45.5} & 36.0 & \underline{45.2} & 26.8 & 88.6 \\
Accuracy (CI) & \textbf{48.7} & 40.4 & \underline{45.6} & 22.9 & 85.9 \\
Accuracy (R) & \textbf{57.3} & 56.2 & \underline{57.2} & 42.6 & 89.7 \\
\arrayrulecolor{darkgray}\midrule\arrayrulecolor{black}
IoU & 52.0 & \textbf{53.8} & \underline{52.1} & 50.9 & 59.9 \\
Exact Match & \underline{17.0} & \textbf{18.1} & 16.2 & 16.3 & 24.0 \\
Macro F1 & 57.1 & \textbf{60.9} & 57.2 & \underline{57.8} & 62.6 \\
\arrayrulecolor{darkgray}\midrule\arrayrulecolor{black}
MAE & \underline{17.7} & \textbf{17.5} & \underline{17.7} & \textbf{17.5} & 16.0 \\
\bottomrule
\end{tabular}

%% file: tables/stito_results.tex
\begin{tabular}{lccc}
\toprule
 & \multicolumn{3}{c}{Condition} \\
\cmidrule(lr){2-4}
Configuration & WS & WI & CI \\
\midrule
$z_T$ & \underline{0.390\,{\scriptsize$\pm$0.010}} & \textbf{0.575\,{\scriptsize$\pm$0.016}} & \textbf{0.647\,{\scriptsize$\pm$0.018}} \\
$z_y$ & \textbf{0.383\,{\scriptsize$\pm$0.010}} & \underline{0.626\,{\scriptsize$\pm$0.019}} & \underline{0.720\,{\scriptsize$\pm$0.021}} \\
\arrayrulecolor{darkgray}\midrule\arrayrulecolor{black}
AFx-Rep \cite{steinmetz2024st} & 0.393\,{\scriptsize$\pm$0.011} & 0.663\,{\scriptsize$\pm$0.018} & 0.832\,{\scriptsize$\pm$0.023} \\
Fx-Enc++ \cite{yeh2025fx} & 0.443\,{\scriptsize$\pm$0.013} & 0.803\,{\scriptsize$\pm$0.027} & 1.144\,{\scriptsize$\pm$0.037} \\
\arrayrulecolor{darkgray}\midrule\arrayrulecolor{black}
CLAP \cite{elizalde2023clap} & 0.539\,{\scriptsize$\pm$0.013} & 0.807\,{\scriptsize$\pm$0.020} & 1.041\,{\scriptsize$\pm$0.028} \\
\bottomrule
\end{tabular}